\documentstyle{lamuphys}

\def\kms{km~s$^{-1}$}
\newcommand{\lsim}{\ \raise -2.truept\hbox{\rlap{\hbox{$\sim$}}\raise5.truept
        \hbox{$<$}\ }}
\newcommand{\gsim}{\ \raise -2.truept\hbox{\rlap{\hbox{$\sim$}}\raise5.truept
        \hbox{$>$}\ }}
\makeatletter
\let\chapter\hid@chapter
\makeatother

\begin{document}
\pagenumbering{arabic}

\input epsf

\title{Impact of SNIa on SED of high redshift galaxies}

\author{E. Brocato\inst{1,3}, S. Savaglio\inst{2}, G. Raimondo\inst{1,3}}

\institute{Osservatorio Astronomico di Collurania, Via M. Maggini,
I--64100 Teramo, Italy
\and
European Southern Observatory, Schwarzschild Str. 2
Garching, D--85748, Germany
\and
Istituto Nazionale di Fisica Nucleare, LNGS, I--67100 L'Aquila,
Italy}

\maketitle

\begin{abstract}
We present preliminary results on the effects of 
SNIa explosions on the Spectral Energy Distribution (SED)
 of distant galaxies and 
the possible modifications which may occur in  integrated
spectra,  magnitudes and colours of simple galaxy models of different 
ages and metallicities few days
after a SNIa event.
\end{abstract}

\section{Introduction}

Recent observations allowed to derive spectra of very high redshift
galaxies ($z \gsim 3$) providing information on their
earlier stage of evolution. Several authors infer galaxy ages
by using the population synthesis technique and ages of the
order of 1 Gyr or less have been suggested.
We are interested in investigating the impact of
SNe on the total emitted light of a galaxy in the range of ages
running from 0.1 Gyr to few Gyr.
The  SNIa events should 
appear for ages comparable to the time scale of the first generation of 
white dwarf (WD) stars ($\simeq 0.05 - 0.1$ Gyr).
We suggest that SNIa may have a non--negligible impact
on the SED of a galaxy. 
We assume few models for the
stellar population synthesis which might be representative of the observed SED of
high redshift galaxies and do not consider obscuration by dust. 
We stress that reproducing all the possible models which explain observed SEDs
is not the goal of this work, while we focus our attention on the
variations of the observable parameters when SNIa 
events are taken into account.

\section{ Photometric impact}
 
The present knowledge of  SN rates is still far from being firm.
At high redshift, 
the effect of the expanding Universe will decrease the SN rate of a
factor $(1+z)$ due to time dilution, 
but the same effect will stretch 
the light curve of a SN by the same factor.
Finally, one  expects to observe a {\it larger} number of SNe Ia
in the young Universe where we should observe the primeval galaxy population.

\begin{figure}
\epsfxsize=10cm
\epsfysize=10cm
\centerline{\epsffile{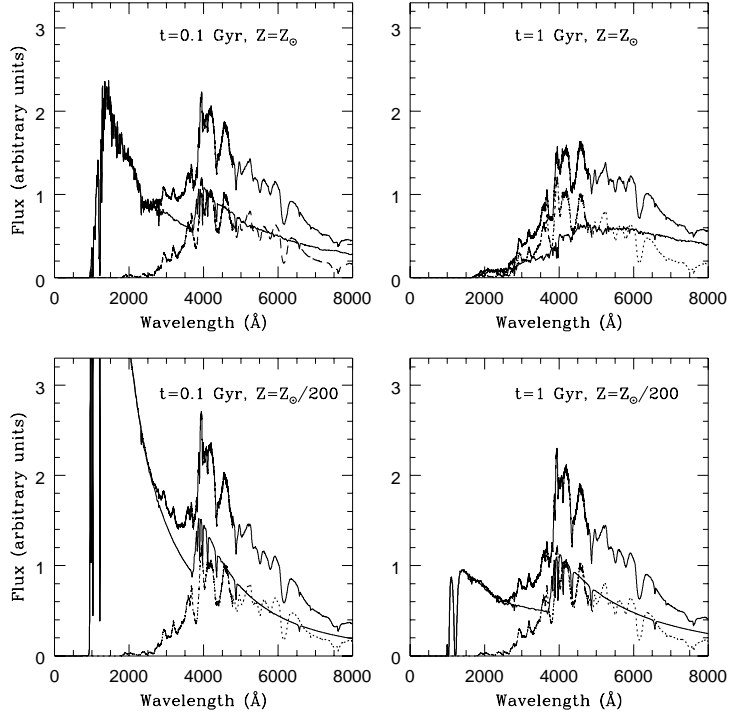}}
\caption[1]{\label{f3}Spectral Energy Distribution for 4 different 
 galaxy models (thin lines). The dotted line is the SNIa spectrum 5 days
 after the explosion normalized to the V magnitude of the
 parent galaxy. Thick lines show the composite spectra 
 (galaxy + SNIa).}
\end{figure}

To evaluate the variation in the SED of the parent galaxy
in case of a SNIa explosion, 
we assume a Simple Stellar Population (SSP) originated by a unique burst
of star formation (Brocato \& Romaniello 1997, in preparation) 
and the stellar atmosphere models by Kurucz 1995 are adopted.
 
The SNIa spectrum refers to SN 1992A (the optical part is kindly
provided by B. Leibundgut whereas the UV part
has been retrieved from the HST archive) and 
it was obtained 5 days after the explosion. 
We simulated a SNIa event in 4 different galaxy models, namely,
$(i)$ $t = 0.1$ Gyr,  $Z=Z_\odot$; $(ii)$
$t = 1.0$ Gyr,  $Z=Z_\odot$; $(iii)$
$t = 0.1$ Gyr,  $Z=Z_\odot/200$; $(iv)$
$t = 1.0$ Gyr,  $Z=Z_\odot/200$.
We assume the V magnitude of the parent galaxy to be equal to that of
the SN at the maximum ($V_{\rm SN}^{max} = -19$ for $H_o=75$ \kms~Mpc$^{-1}$, 
Wheeler \& Benetti, 1996).
The results are shown in Fig.~\ref{f3}
where we plot the SNIa spectrum, the SED of the parent galaxy
and the composite spectrum (galaxy + SNIa).
The some strong spectral features of the SNIa
can be clearly seen in large details making
the SNIa detection 
in distant galaxies a possible task. A rather
interesting absorption  is the SiII$\lambda\lambda 5972,6355$
doublet.

\section{Moving toward high redshift galaxies}

\begin{figure}
\epsfxsize=9.1cm
\epsfysize=9.1cm
\centerline{\epsffile{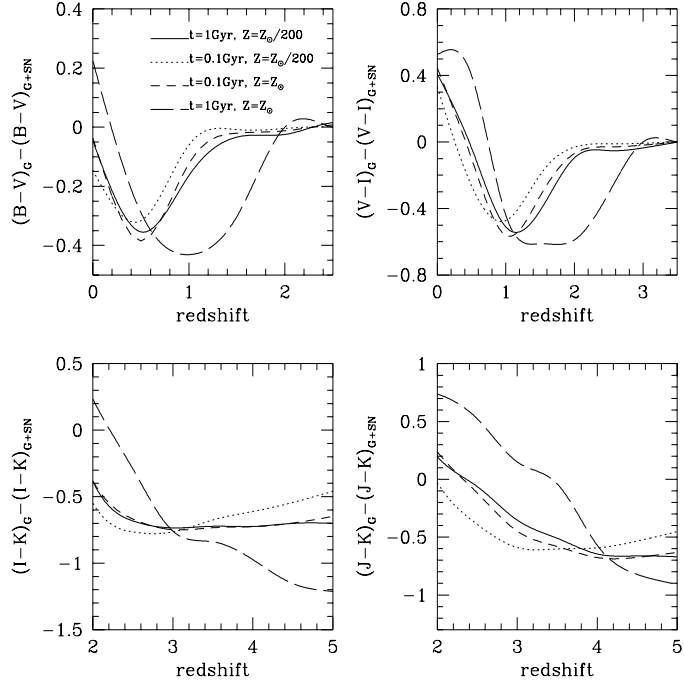}}
\caption[1]{\label{f6}Colour variations for 4 galaxy models due
to the SNIa contribution.}
\end{figure}

We computed some significant colours for 4 galaxy models and plot the
differences in these colours when the SNIa is included (Fig. \ref{f6}). 
With the exception of the galaxy with $t = 1$ Gyr and solar metallicity, the
remaining 3 models have the same behaviour. The J--K colour shows
significant differences at high redshift, being around 0.5 at
$z>3$. The difference is even higher in the I--K colour.
In the optical, the colour differences are significant only at
low redshift ($z<1$).

One single SNIa event can dramatically
change the SED of young galaxies. In general, 
for each given redshift there is a preferred
colour in which the magnitude variation due to the SNIa event is larger.
On the other hand, the SN rates at high redshift
have never been tested. Simple considerations on the 
theoretical expectations and on the galaxy density at high redshift
may allow the probability of finding SN events to be determined and to
be compared with observations. This will also depend on the
morphological galaxy type while the absorption features will be function
of the galaxy metallicity.

\end{document}